\documentclass[12pt]{iopart}
\usepackage{epsfig}
\begin{document}

\title[Nonlinear resonance in a three-terminal carbon nanotube resonator]{Nonlinear resonance in a three-terminal carbon nanotube resonator}

\author{A. Isacsson$^1$ and J. M. Kinaret$^1$}
\address{$^1$Department of Applied Physics, Chalmers University of Technology, SE-412 96 G\"oteborg, Sweden}
\ead{andreas.isacsson@chalmers.se}
\author{R. Kaunisto$^2$}
\address{$^2$Nokia Research Center, P.O. Box 407, FIN-00045 Nokia Group, Helsinki, Finland}
\begin{abstract}
The RF-response of a three-terminal carbon nanotube resonator coupled to RF-transmission lines is studied by means of perturbation theory and direct numerical integration. 
We find three distinct oscillatory regimes, including one regime capable of exhibiting very large hysteresis loops in the frequency response. Considering a purely capacitive transduction, we derive a set of algebraic equations which can be used to find the output power (S-parameters) for a device connected to transmission lines with characteristic impedance $Z_0$.  
\end{abstract}

\pacs{00.00, 00.00, 00.00}
\submitto{\NT}
\maketitle

\section{Introduction}
Nanoelectromechanical systems (NEMS) are systems where mechanical and electronic degrees of freedom are coupled and whose characteristic length
scales are measured in nanometers. While metal or silicon are common material choices for microelectromechanical devices (MEMS), carbon nanotubes (CNT) may become one of the mainstays in future NEMS technology~\cite{Xu:2003} due to their unique combination of electrical and mechanical properties: small mass, extraordinary stiffness, low mechanical dissipation and electrical properties 
ranging from semiconducting to conducting~\cite{CNTPROPERTIES}. Combined together, these properties allow for NEMS devices operating in the high GHz regime.
Several CNT-based NEMS devices have already been demonstrated~\cite{Sazanova:2004,CNTNEMS2,Herre:2006}.

\begin{figure}[t]
\begin{center}
\epsfig{file=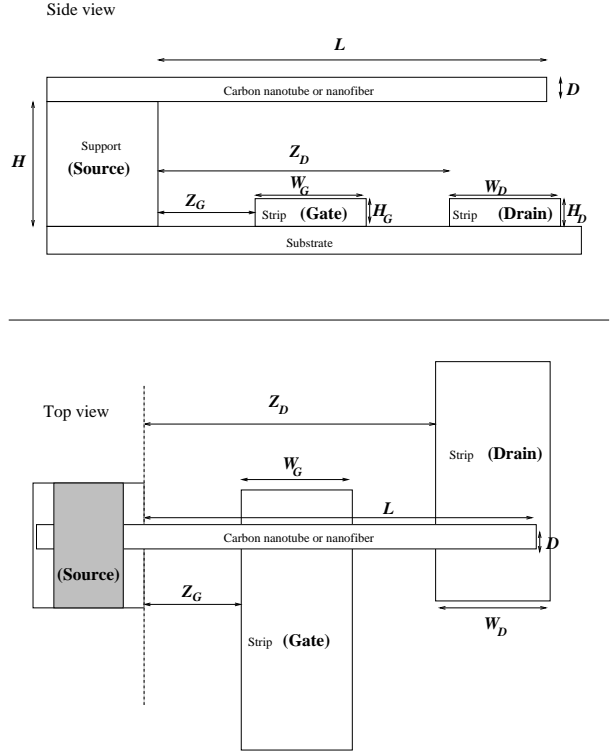, width=8cm,clip}\end{center}
\caption{Physical layout of a three-terminal carbon nanotube resonator. A carbon nanotube or carbon fiber with diameter $D$ is mounted on a support -- the source contact -- of height $H$ extruding a distance $L$ out over the substrate. Beneath the tube are two metallic strips -- the Gate and the Drain -- of widths $W_{\rm G}$ and $W_{\rm D}$ respectively. The distances between the gate and drain strips from the support are $Z_{\rm G}$ and $Z_{\rm D}$ and the strips have heights $H_{\rm G}$ and $H_{\rm D}$ respectively. \label{fig:1} }
\end{figure}

In \cite{Kinaret1,Kinaret2,Hwang:2005} one specific such carbon-nanotube-based system was considered: a three-terminal nanomechanical relay with a layout 
similar the one in figure~\ref{fig:1}. Such devices have since been successfully fabricated~\cite{Lee:2004, Axelsson:2005,SACLAY}. In \cite{Kinaret1,Kinaret2,Lee:2004, Axelsson:2005,SACLAY} the device was considered mainly operating as a switch and transduction was based on a 
tunneling current between the tube tip and the drain terminal. In a subsequent publication~\cite{Kinaret3} also the high frequency properties were considered 
and nonlinear resonant behavior was demonstrated. 

Nonlinear response to AC-drive is a characteristic feature of many MEMS devices as well as in NEMS~\cite{CLELANDNEMSBOOK} and has recently received much attention~\cite{Zaitsev:2005,Aldridge:2005, Postma:2005, Kozinsky:2005,Almog:2006_1,Almog:2006_2}. It can typically be modeled by using the Duffing equation. Understanding of the parametric dependence on the behavior of this response is essential for any potential technological application. Of equal importance for technological applications is to understand how NEMS devices act when embedded in an electronic circuit. In this paper we investigate more carefully the RF-response of the three-terminal CNT resonator in figure~\ref{fig:1} subjected to a harmonic input signal on the gate electrode, going beyond the simple Duffing equation. In contrast to previous publications~\cite{Kinaret3}, we base the transduction not on electron tunneling between the tube tip and the drain but on the displacement current generated by the time-varying tube-drain capacitance when the device is connected to lossless transmission lines.

This paper consists of several sections. In section 2 we present the system and derive a lumped dynamical model. Then, by means of perturbation theory, we derive in section 3 a set of algebraic equations that can be efficiently used to obtain and classify the frequency response of the device along with expressions for the output power ($S_{21}$-parameter). In section 4 we consider the regime of linear response solutions and in section 5 we characterize the nonlinear behavior. Finally in section 6 we discuss the domain of validity of the perturbative approach by direct comparison with numerical integration and discuss the delivered power arising from a purely capacitive transduction.

\section{Model}
\label{Sec:Model}

The system is depicted in figure~\ref{fig:1}. Mounted on a conducting support (Source) of height $H$ is a nanotube of length $L$ with outer diameter $D_{\rm o}$ and an inner diameter $D_{\rm i}$. Below the tube two conducting strips of heights $H_{{\rm G,D}}$ and widths $W_{{\rm G,D}}$ act as gate and drain electrodes. 
The motion of the carbon nanotube can be described as a simple elastic beam deflecting in only one direction~\cite{Yakobson} using the action 
\begin{equation*}
S_{\rm beam}=\int_{t_1}^{t_2} dt \int_0^L dx \frac{\rho A}{2}[\partial_t u(x,t)]^2-\frac{EI}{2}[\partial_x^2 u(x,t)]^2.
\end{equation*}
Here $u(x,t)$ is the instantaneous deviation of the tube towards the Drain electrode, at a distance $x$ from the tube support.
$E$ is the effective Young modulus\cite{Yakobson, Wong:1997, Poncharal:1999} of the beam and $I=\pi (D_{\rm o}^4-D_{\rm i}^4)/64$ the moment of inertia. The cross-section area is $A=\frac{1}{4}\pi (D_{\rm o}^2-D_{\rm i}^2)$ 
and $\rho$ is the density of the tube.
The external forces acting on the tube, i.e. actuation and transduction, arise from 
capacitive coupling between the tube and the electrodes. These forces depend on the instantaneous charge distribution and geometrical configuration of the tube. 
This gives rise to an additional part of the action for the tube which can be written
\begin{equation*}
S_{\rm el.-mech.}=-\int_{t_1}^{t_2} dt \frac{1}{2}\sum_{i={\rm G,D}}\frac{Q^\prime_i(t)^2}{C_i[u(\cdot,\cdot),t]}
\end{equation*}
where $C_{{\rm G,D}}[u(x,t),t]$ are the tube-gate and tube-drain capacitances at time $t$ and $Q^\prime_{{\rm G,D}}$ the associated capacitor charges.
Stationarity of the action with respect to $u(x,t)$ then provides the equation of motion for the beam
\begin{equation}
\rho A\partial_t^2 u(x,t)+{EI}\partial_x^4 u(x,t)=\frac{1}{2}\sum_{i={\rm G,D}}[\Delta V_i(t)]^2\frac{\delta C_i[u,t]}{\delta u(x)}
\label{eq:pde}
\end{equation}
where we have introduced the shorthand
\begin{equation*}
\frac{\delta C_i[u,t]}{\delta u(x)}=\int dt' \frac{\delta C_i[u,t]}{\delta u(x,t')}
\end{equation*}
to emphasize that the functional derivative only affects the spatial dependence of the capacitance at time $t$ and the potential differences $\Delta V_{{\rm G,D}}=V_{\rm T}-V_{{\rm G,D}}$ between the tube and the gate/drain electrodes.
\begin{figure}
\begin{center}
\epsfig{file=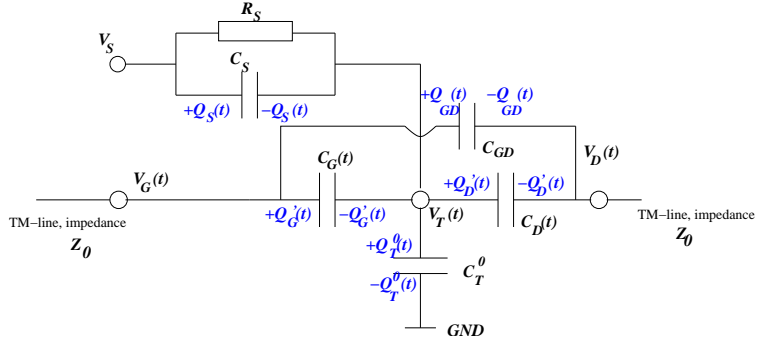, width=10cm, clip}\end{center}
\caption{Equivalent circuit model. The gate-tube capacitance $C_{\rm G}$ as well as the drain-tube capacitance $C_{\rm D}$ depend on the instantaneous deflection of the cantilever. The gate and drain contacts are assumed to be connected to transmission lines with characteristic impedance $Z_0$. \label{fig:2}}
\end{figure}

The electronic degrees of freedom are most conveniently treated within the circuit model in figure~\ref{fig:2}. 
For finite source-tube resistance $R_{\rm S}$ we have three relevant degrees of freedom $Q_{\rm G}(t)\equiv Q'_{\rm G}(t)+Q_{\rm GD}(t)$,
$Q_{\rm D}(t)\equiv Q'_{\rm D}(t)+Q_{\rm GD}(t)$ and $Q_{\rm T}(t)\equiv Q'_{\rm D}(t)+Q_{\rm T}^0(t)-Q'_{\rm G}(t)-Q_{\rm S}(t)$.
With only one incoming signal on the gate, $V_+(t)$, these charges obey the equations of motion 
\begin{eqnarray}
\partial_tQ_{\rm G}=\frac{1}{Z_0}\left[2V_{+}(t)-V_{\rm G}(t)+V_{\rm G}^0\right]\label{eq:QDE1}\\
\partial_tQ_{\rm D}=\frac{1}{Z_0}\left[V_{\rm D}(t)-V_{\rm D}^0\right]\label{eq:QDE2}\\
\partial_tQ_{\rm T}=\frac{1}{R_{\rm S}}[V_{\rm S}-V_{\rm T}(t)]\label{eq:QDE3},
\end{eqnarray}
where we have allowed for constant DC-bias offsets on each of the electrodes and assumed the gate and drain connected to lossless transmission lines with a real impedance $Z_0$. For the tube diameters and signal frequencies we consider, the kinetic inductance of the CNT~\cite{Burke:2003,Zhang:2006} is insignificant in relation to other impedances. 
The capacitances give us a linear relation ship between voltages $(V_{\rm G},V_{\rm D},V_{\rm T})$ and charges $(Q_{\rm G},Q_{\rm D},Q_{\rm T})$,
\begin{eqnarray}
 Q_{\rm G}=(C_{\rm G}+C_{\rm GD})V_{\rm G}-C_{\rm G}V_{\rm T}-C_{\rm GD}V_{\rm D}\label{eq:QV1}\\
 Q_{\rm D}=C_{\rm D}V_{\rm T}+C_{\rm GD}V_{\rm G}-(C_{\rm D}+C_{\rm GD})V_{\rm D}\label{eq:QV2}\\
 Q_{\rm T}=(C_{\rm T}^0+C_{\rm S}+C_{\rm D}+C_{\rm G})V_{\rm T}-C_{\rm S}V_{\rm S}-C_{\rm D}V_{\rm D}-C_{\rm G}V_{\rm G}\label{eq:QV3}
\end{eqnarray} 
ensuring that we have a closed set of equations for the dynamics consisting of equations~(\ref{eq:pde})-(\ref{eq:QV3}).

The full PDE in (\ref{eq:pde}) is not very tractable as it stands. We will assume that only the 
lowest lying fundamental vibration mode is excited by the incoming signal. In this approximation  the deformation of the tube can be parameterized by the displacement $x_{\rm T}$ of the cantilever tip from the static equilibrium position (for details see Appendix A)
\begin{equation}
{\ddot{x}_{\rm T}(t)+\Gamma \dot{x}_{\rm T}+\Omega_0^2x_{\rm T}(t)}=\frac{1}{2M_{\rm eff}}\sum_{i={\rm G,D}}[\Delta V_i(t)]^2C_i'(x_{\rm T}).
\label{eq:lumpad}
\end{equation}
where $M_{\rm eff}=\rho A L/5.684$ and $\Omega_0={3.516}L^{-2}\sqrt{EI/\rho A}$.
Also, in (\ref{eq:lumpad}) a factor $\Gamma \dot{x}_{\rm T}$ has been incorporated in a phenomenological way to account for dissipation~\cite{Poncharal:1999,Jiang:2004}.

\section{Perturbation theory}
Although direct numerical integration of the lumped model is a straightforward task it is time consuming due to the multiple time scales involved. It is instead our aim in this paper to derive a set of algebraic equations which can be used to classify or to quickly determine the response of a given geometric or biasing configuration.   
We do this through a perturbative analysis by means of the averaging method~\cite{Nayfeh}. We will thus assume that we have a single incoming signal on the Gate electrode $V_+(t)=V_+\cos(\Omega t)$ and cast the system into a dimensionless form by writing $t=\tau/\Omega_0$, $x_{\rm T}=\xi H$, $C_i=c_iC_0$, $V_i=v_iV_0$, $Q_i=q_iV_0C_0$, $\gamma=\Gamma/\Omega_0$ and $\omega=\Omega/\Omega_0$; 
\begin{eqnarray}
 \ddot{\xi}+\gamma\dot{\xi}_{\rm T}+\xi=\frac{\sigma}{2}[v_{\rm G}(\tau)-v_{\rm T}(\tau)]^2c_{\rm G}'(\xi)+\frac{\sigma}{2}[v_{\rm D}(\tau)-v_{\rm T}(\tau)]^2c_{\rm D}'(\xi)\nonumber\\
 \dot{q}_{\rm G}(\tau)=\epsilon^{-1}\left[2v_+(\tau)\cos \omega\tau-v_{\rm G}(\tau)+v_{\rm G}^0\right]\nonumber\\
 \dot{q}_{\rm D}(\tau)=\epsilon^{-1}\left[v_{\rm D}(\tau)-v_{\rm D}^0\right]\nonumber\\
 \dot{q}_{\rm T}(\tau)=\epsilon^{-1}\frac{Z_0}{R_{\rm S}}\left[v_{\rm S}-v_{\rm T}(\tau)\right].\nonumber
\end{eqnarray}
Here we have defined $\sigma\equiv{V_0^2C_0}/M_{\rm eff}\Omega_0^2H^2={\cal E_C}/{\cal E_K}$ and $\epsilon\equiv {\Omega_0 Z_0C_0}$.
For $C_0$ we chose $C_0\equiv \sqrt{C_{\rm G}^0C_{\rm D}^0}$ where $C_{{\rm G,D}}^0$ are the gate-tube and drain-tube capacitances when the tube is undeflected. 
Typically the capacitances in nanoscale devices lie in the attofarad range which implies that for a single device the impedance mismatch is large. Hence, $\epsilon \ll 1$ providing a good starting point for perturbation theory. The linear relationships between charges and voltages imply that we can expand in  $\epsilon$ 
\begin{equation}
q_i=\sum_{n=0}^\infty \epsilon^n q_i^{(n)}(\tau),\quad v_i=\sum_{n=0}^\infty \epsilon^n v_i^{(n)}(\tau), i={\rm G,D,T}.
\label{eq:pertexp}
\end{equation}
To zeroth order in $\epsilon$ we then have
\begin{eqnarray}
v_{\rm G}^{(0)}(\tau)=2v_+\cos \omega\tau+v_{\rm G}^0\nonumber\\
v_{\rm D}^{(0)}(\tau)=v_{\rm D}^0,\quad v_{\rm T}^{(0)}(\tau)=v_{\rm S}.\nonumber
\end{eqnarray}
Inserting these zeroth order solutions into the dynamic equation for the tip motion we find
\begin{eqnarray}
 \ddot{\xi}+\gamma\dot{\xi}_{\rm T}+\xi=\frac{\sigma}{2}[2v_+\cos\omega \tau-(v_{\rm S}-v_{\rm G}^0)]^2c_{\rm G}'(\xi)+\frac{\sigma}{2}[v_{\rm D}^0-v_{\rm S}]^2c_{\rm D}'(\xi)\nonumber.
\end{eqnarray}
We now make the Ansatz of an oscillatory solution with slowly changing parameters, 
\begin{equation*}
\xi=x_0(\tau)+r(\tau)\cos[\omega \tau +\phi(\tau)],
\end{equation*}
and assume that the quantities ${\dot{r}}/{\omega r},{\dot{\phi}}/{\omega \phi}\ll 1$ keeping only the lowest order terms 
\begin{eqnarray}
\gamma\dot{x}_0&+&x_0-(2\dot{r}\omega+\gamma r\omega)\sin(\omega\tau+\phi)\nonumber\\
&+&r(1-\omega^2-2\omega\dot{\phi})\cos(\omega\tau+\phi)=\frac{\sigma}{2}K(x_0,r,\phi,\tau).\nonumber
\end{eqnarray}
Here the kernel $K(x_0,r,\phi,\tau)$ is defined as
\begin{eqnarray}
K&\equiv& [2v_+\cos\omega\tau-(v_{\rm S}-v_{\rm G}^0)]^2c_{\rm G}'(x_0+r\cos[\omega\tau+\phi])\nonumber\\
&+&(v_{\rm D}^0-v_{\rm S})^2c_{\rm D}'(x_0+r\cos[\omega\tau+\phi]).\nonumber
\end{eqnarray}
Provided the slow variables $r,\phi$ and $x_0$ do not change appreciably during one period we can average over one period to find
\begin{eqnarray}
&&\gamma\dot{x}_0+x_0=\frac{\sigma\omega}{4\pi}\int_0^{\omega/2\pi}d\tau\, K(x_0,r,\phi,\tau)\nonumber\\
&&-\dot{r}\omega-\frac{1}{2}\gamma r\omega=\frac{\sigma\omega}{4\pi}\int_0^{\omega/2\pi}d\tau\, K\sin(\omega\tau+\phi)\nonumber\\
&&\frac{1}{2}r(1-\omega^2)-r\omega\dot{\phi}=\frac{\sigma\omega}{4\pi}\int_0^{\omega/2\pi}d\tau\, K\cos(\omega\tau+\phi).\nonumber
\end{eqnarray}
For generic capacitances we would not expect to be able to do these integrals exactly. However, for the case when $z_{\rm D}+w_{\rm D}/2<l$ the derivative of the capacitances can be approximated by an inverse square law according to the local tube deflection,
$c_{{\rm G,D}}'(\xi)={c_{{\rm G,D}}^0\alpha_{{\rm G,D}}}{(\alpha_{{\rm G,D}}-\xi)^{-2}}$
with 
\begin{equation*}
\alpha_{{\rm G,D}}\equiv(1-H_{{\rm G,D}}/H){u_0(L)}/{u_0({Z_{{\rm G,D}}/L+W_{{\rm G,D}}/2L})}.
\end{equation*}
Introducing 
\begin{equation*}
a_{{\rm G,D}}\equiv\frac{1}{r}\left(\alpha_{{\rm G,D}}-x_0+\sqrt{(\alpha_{{\rm G,D}}-x_0)^2-r^2}\right)
\end{equation*} 
and
$v_{\rm SG}\equiv v_{\rm S}-v_{\rm G}^0$, $v_{\rm SD}\equiv v_{\rm S}-v_{\rm D}^0$ and $\tilde{v}_{\rm SG}^2\equiv{2v_+^2+v_{\rm SG}^2}$,
the results after performing the integrals are
\begin{eqnarray}
\fl \gamma\dot{x}_0+x_0=\frac{2\sigma c_{\rm G}^0\alpha_{\rm G}a_{\rm G}^2}{r^2(a_{\rm G}^2-1)^3}[\tilde{v}_{\rm SG}^2(a_{\rm G}^2+1)- 8v_+v_{\rm SG}a_{\rm G}\cos\phi+2v_+^2(3-a_{\rm G}^{-2})\cos 2\phi]\nonumber \\
+\frac{2\sigma c_{\rm D}^0\alpha_{\rm D}a_{\rm D}^2(a_{\rm D}^2+1)}{r^2(a_{\rm D}^2-1)^3}v_{\rm SD}^2\label{eq:p1}\\
\fl 2\dot{r}\omega+\gamma r\omega=8\sigma\frac{c_{\rm G}^0\alpha_{\rm G}}{r^2(a_{\rm G}^2-1)}\left[v_+v_{\rm SG}\sin\phi-v_+^2a_{\rm G}^{-1}\sin 2\phi\right]\label{eq:p2}\\
\fl r(1-\omega^2)-2r\omega\dot{\phi}=8\sigma\frac{c_{\rm G}^0\alpha_{\rm G}a_{\rm G}^2}{r^2(a_{\rm G}^2-1)^3}[\tilde{v}_{\rm SG}^2a_{\rm G}-v_+v_{\rm SG}(a_{\rm G}^2+4-a_{\rm G}^{-2})\cos\phi\nonumber \\
+v_+^2(a_{\rm G}+2a_{\rm G}^{-1}-a_{\rm G}^{-3})\cos2\phi]+8\sigma\frac{c_{\rm D}^0\alpha_{\rm D}a_{\rm D}^3}{r^2(a_{\rm D}^2-1)^3}v_{\rm SD}^2\label{eq:p3}
\end{eqnarray}

The output signal is the displacement current in the drain contact and we will express this in terms of the 
square modulus of the S-parameter $|S_{21}|^2$. In general the S-parameter $S_{ij}$ is a complex quantity relating the amplitude and phase of an incoming signal $V_i$ on port $i$ to an outgoing signal on port $j$ through $S_{ij}\equiv V_i/V_j$. In terms of power we have then  $|S_{21}|^2={\bar{P}_{\rm out}}/{\bar{P}_{\rm in}}$ relating the total outgoing RF-power on the drain to the incoming RF-power on the gate. 
For the given input signal $v_+\cos(\omega \tau)$ the average delivered power to the device is
$\bar{P}_{\rm in}={V_0^2v_+^2}/{2Z_0}$
and the total power delivered on the output is  
\begin{equation*}
\bar{P}_{out}=\frac{V_0^2\omega}{2Z_0\pi}\int_0^{\frac{\omega}{2\pi}}d\tau (v_{\rm D}^0-v_{\rm D}(\tau))^2.
\end{equation*}
From the linear relation between charges and voltages we have
\begin{equation*}
q_{\rm D}(\tau)=c_{\rm D}v_{\rm T}+c_{\rm GD}v_{\rm G}-(c_{\rm D}+c_{\rm GD})v_{\rm D}.
\end{equation*}
Recalling the perturbation expansion~(\ref{eq:pertexp})
we get to lowest order in $\epsilon$
\begin{equation*}
v_{\rm D}=v_{\rm D}^0+\epsilon((v_{\rm S}-v_{\rm D}^0)c_{\rm D}'(\xi)\dot{\xi}-2\omega v_+c_{\rm GD}\sin\omega\tau)+O(\epsilon^2).
\end{equation*}
Evaluating the integral in the stationary state ($\dot{x}_0=\dot{\phi}=\dot{r}=0$) one finds
\begin{eqnarray}
|S_{21}|^2&=&4(Z_0C_0\Omega)^2\left[4\left(\frac{v_s-v_{\rm D}^0}{v_+}\right)^2\frac{(c_{\rm D}^0)^2\alpha_{\rm D}^2a_{\rm D}^4(a_{\rm D}^2+1)}{r^2(a_{\rm D}^2-1)^5}\right.\nonumber\\
&+&\left.4(\frac{v_{\rm S}-v_{\rm D}^0}{v_+})\frac{c_{\rm D}^0c_{\rm GD}\alpha_{\rm D}}{r(a_{\rm D}^2-1)}\cos\phi+c_{\rm GD}^2\right].\label{eq:S21}
\end{eqnarray}
A similar result can be derived for the reflection coefficient $|S_{11}|^2$. Note that the prefactor of $\epsilon^2=(Z_0C_0\Omega)^2$ indicates that for mismatched systems 
($\epsilon\ll 1$) only a very small amount of the incoming power is actually delivered to the device and that most is reflected back ($|S_{11}|^2\sim 1$).

In a typical setup we have $\alpha_{\rm G}\gg\alpha_{\rm D}$. In this case we have found that the system (\ref{eq:p1})-(\ref{eq:p3}) can be simplified considerably by omitting all terms related to the parametric driving and double frequency components and it suffices to solve the simplified set of equations
\begin{eqnarray}
x_0&=&\frac{\sigma c_{\rm G}^0}{2\alpha_{\rm G}}\tilde{v}_{\rm SG}^2+\frac{2\sigma c_{\rm D}^0\alpha_{\rm D}a_{\rm D}^2(a_{\rm D}^2+1)}{r^2(a_{\rm D}^2-1)^3}v_{\rm SD}^2\label{eq:p21}\\
\gamma r\omega&=&2\frac{\sigma c_{\rm G}^0}{\alpha_{\rm G}}v_+v_{\rm SG}\sin\phi\label{eq:p22}\\
(1-\omega^2)&=&\frac{\sigma c_{\rm G}^0}{\alpha_{\rm G}^2}\tilde{v}_{\rm SG}^2-2\frac{\sigma c_{\rm G}^0}{r\alpha_{\rm G}^2}v_+v_{\rm SG}\alpha_{\rm G}\cos\phi\nonumber\\
&+&8\sigma\frac{c_{\rm D}^0\alpha_{\rm D}a_{\rm D}^3}{r^3(a_{\rm D}^2-1)^3}v_{\rm SD}^2\label{eq:p23}.
\end{eqnarray}
Solving this simplified system produces results which agree quantitatively with the full system~(\ref{eq:p1})-(\ref{eq:p3}) for weak driving and qualitatively for all bias ranges.

\section{Static solutions and linear response}
We consider first the statics and small amplitude vibrations around equilibrium and derive a formula for the output power in this regime. 
To this end we expand (\ref{eq:p1})-(\ref{eq:p3}) to first order
in  $v_+/(v_{\rm S}-v_{\rm G}^0)$ and $r/(\alpha_{{\rm G,D}}-x_0)$, i.e., we assume the oscillation amplitude to be small compared to the maximum amplitude allowed for a given static deflection
and obtain
\begin{eqnarray}
\gamma r\omega&=&2\sigma\frac{c_{\rm G}^0\alpha_{\rm G}v_+v_{\rm SG}}{(\alpha_{\rm G}-x_0)^2}\sin\phi\nonumber\\
(\omega^2-\omega_r^2)&=&2\sigma\frac{c_{\rm G}^0\alpha_{\rm G}v_+v_{\rm SG}}{r(\alpha_{\rm G}-x_0)^2}\cos\phi\nonumber.
\end{eqnarray}
Here $x_0$ is the deflection in the absence of drive, i.e., $v_+=0$ 
\begin{equation*}
x_0=\frac{\sigma}{2}\left[\frac{c_{\rm G}^0\alpha_{\rm G} v_{\rm SG}^2}{(\alpha_{\rm G}-x_0)^2}+\frac{c_{\rm D}^0\alpha_{\rm D}v_{\rm SD}^2}{(\alpha_{\rm D}-x_0)^2}\right]
\end{equation*}
and $\omega_r$ is the renormalized frequency
\begin{equation*}
\omega_r^2=1-\frac{\sigma}{2}\left[\frac{c_{\rm G}^0\alpha_{\rm G}v_{\rm SG}^2}{(\alpha_{\rm G}-x_0)^3}+\frac{c_{\rm D}^0\alpha_{\rm D}v_{\rm SD}^2}{(\alpha_{\rm D}-x_0)^3} \right].
\end{equation*}

\begin{figure}[t]
\begin{center}
\epsfig{file=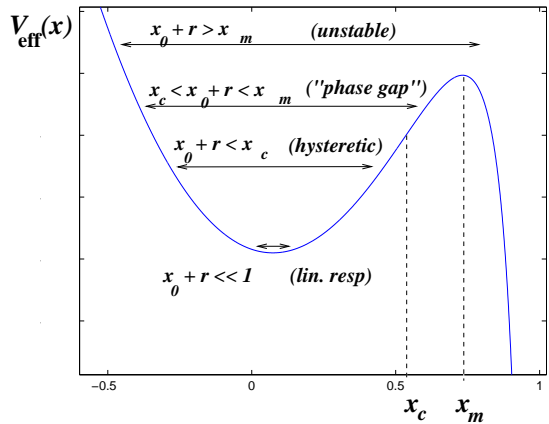, width=8cm,clip}
\end{center}
\caption{Typical shape of effective potential for the driven oscillator. Marked in the figure are the maximum $x_{\rm m}$ above which the system is unstable towards snap-to-contact and the locus of the point where the second derivative changes sign, $V_{\rm eff}^{\prime\prime}(x_{\rm c})=0$. Also shown are the typical oscillation regimes and their amplitudes. The qualititave difference between these regimes is best understood from looking at the phase $\phi$ of the oscillator relative to the driving field (see figure~\ref{fig:phi}).\label{fig:pot}}
\end{figure}

We note that while the equation for $x_0$ can have two solutions only the solution with the smaller deflection is stable. The solution with large deflection is always unstable and leads to snap to contact. If surface forces are taken into account or if the drain electrode is placed outside the reach of tip 
bistable operation can be obtained as in \cite{Kinaret3}. From the above relations we find in the linear response regime a power transmission of
\begin{eqnarray}
|S_{21}|^2&=&4\epsilon^2\left[\sigma^2\kappa_{\rm G}^2\kappa_{\rm D}^2\frac{\omega^2}{(\omega^2-\omega_r^2)^2+\gamma^2\omega^2}+\omega^2c_{\rm GD}^2\right. \nonumber\\
&+&\left.2\sigma{\kappa_{\rm G}\kappa_{\rm D} c_{\rm GD}}\frac{\omega^2(\omega^2-\omega_r^2)}{(\omega^2-\omega_r^2)^2+\gamma^2\omega^2}\right]
\end{eqnarray}
with $\kappa_{{\rm G,D}}\equiv c_{{\rm G,D}}\alpha_{{\rm G,D}}(v_{\rm S}-v_{{\rm G,D}}^0)/(\alpha_{{\rm G,D}}-x_0)^2$.

\section{Nonlinear response}
We now go beyond the linear response regime and consider the full solutions of (\ref{eq:p1})-(\ref{eq:p3}).  For the general case these equations need to be solved numerically. In order to illustrate the typical resonant behaviour we will consider a specific system with a multi-walled carbon nanotube, $D_{\rm o}=30$ nm, $D_{\rm i}=20$ nm extending a length $L=250$ nm out from the support. The electrode dimensions are (see figure~\ref{fig:1}) $Z_{\rm G}=100$ nm, $Z_{\rm D}=225$ nm, $W_{\rm D}=W_{\rm D}=50$ nm, $H=25$ nm and $H_{\rm G}=H_{\rm D}=10$ nm. Finite element modeling of this structure gives us the capacitances for an unbent configuration $C_{\rm G}^0=5.4$ aF, $C_{\rm D}^0=4.1$ aF, $C_{\rm GD}=6.1$ aF and $C_{\rm T}^0=4$ aF. For the mechanical properties of the nanotubes we have assumed an effective Young modulus of $E=1$ TPa and a quality factor of $Q=200$~\cite{Salvetat:1999,Tombler:2000,Kim:2002,Demczyk:2002,Li:2003,Liu:2001,Liu:2003,Wang:2004}.
\begin{figure}[t]
\begin{center}
\epsfig{file=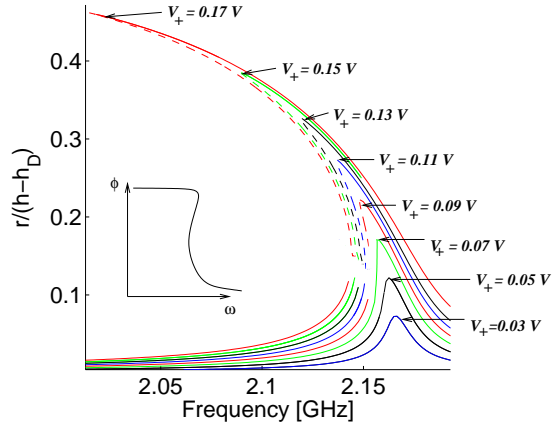, width=8cm,clip}
\end{center}
\caption{Amplitude of oscillation in linear response ($V_+=0.03,...,0.07$ V) and hysteretic regime ($V_+=0.09,...,0.17$ V). (\full) stable solutions, (\dashed) unstable solutions.  \label{fig:r} }
\end{figure}
 
For a qualitative understanding of the response it is useful to look at the non-averaged equations of motion.
If we ignore excitations of other than the fundamental frequency we have the differential equation
\begin{equation*}
\ddot{\xi}+\gamma\dot{\xi}_{\rm T}+\frac{\rmd}{\rmd\xi}V_{\rm eff}(\xi)={\sigma}v_+v_{\rm SG}c_{\rm G}'(\xi)\cos \omega\tau
\end{equation*}
with $V_{\rm eff}=\frac{1}{2}\left[\xi^2-\sigma v_{\rm SD}^2c_{\rm D}(\xi)-\sigma v_{\rm SG}^2c_{\rm G}(\xi)\right].$
At a Source bias $V_{\rm S}=5$ V and with $V_{\rm D}^0=V_{\rm G}^0=0$ V this potential has the shape as shown in figure~\ref{fig:pot} and we can clearly distinguish a few different scenarios. For small excitations we expect to obtain the linear response solution stated above. As driving and amplitude increase we reach a point where deviations from the parabolic potential become manifest, leading to a hysteresis downward in the frequency plane. We expect this regime to have a similar frequency response as the Duffing oscillator. This regime is discussed below in subsection {\it 5.1}.
 As one increases the driving further the oscillation amplitude is expected to increase. As it increases above the point where the curvature of the effective potential changes sign there will be a qualitative change in the response. This change is most markedly seen in the phase response (see figure~\ref{fig:phi}) as ``gap'' around $\phi=\pi/2$. We will discuss this regime in subsection {\it 5.2}.

\subsection{Onset of hysteresis}
\begin{figure}[t]
\begin{center}
\epsfig{file=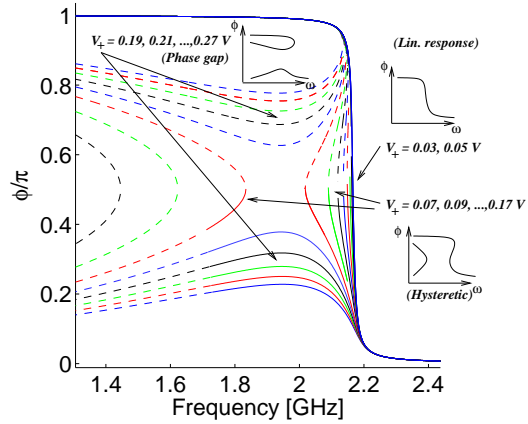, width=8cm,clip}
\end{center}
\caption{Relative phase of oscillations for varying AC-drive amplitude $V_+$. Solid lines indicate stable solutions to (\ref{eq:p1})-(\ref{eq:p3}) while dashed lines correspond to unstable solutions. For small drive amplitudes $V_+\le 0.05$ V the system response agrees with linear response and only stable solutions exist. For intermediate values ($V_+=0.07,...,0.17$ V) hysteresis in the frequency plane appears along high amplitude solutions (mainly unstable). Increasing drive amplitude further a gap in the phase response curve opens up around $\phi=\pi/2$ (curves $V_+=0.19,...,0.27$ V). (\full) stable solutions, (\dashed) unstable solutions.  \label{fig:phi} }
\end{figure}

For sufficiently small vibrations (\ref{eq:p1})-(\ref{eq:p3}) can be expanded in terms of $x_0$ and $r$ to obtain 
the frequency response equation
\begin{eqnarray}
{(\gamma\omega)^2}r^2&+&r^2\left[(\tilde{\omega}^2-\omega^2)-\beta^2 r^2\right]^2=4\left(\frac{\sigma c_{\rm G}^0}{\alpha_{\rm G}}\right)^2v_+^2 v_{\rm SG}^2\label{eq:fresp}
\end{eqnarray}
where
\begin{equation*}
\beta^2\equiv \frac{3\sigma c_{\rm D}^0}{4\alpha_{\rm D}^4} v_{\rm SD}^2
\frac{{\sigma c_{\rm D}^0} v_{\rm SD}^2+2\alpha_{\rm D}^2}{\alpha_{\rm D}^2-{\sigma c_{\rm D}^0} v_{\rm SD}^2}
\end{equation*}
and
\begin{equation*}
\tilde{\omega}^2\approx  1-\frac{\sigma c_{\rm G}^0}{\alpha_{\rm G}^2}\tilde{v}_{\rm SG}^2-\frac{2}{3}\beta^2\alpha_{\rm D}^2.
\end{equation*}
This third order equation is the same as for the Duffing equation and is adequate for determining the onset of the hysteretic behavior. 
In the present case it performs less well to determine the frequency response for intermediate oscillation amplitudes in the hysteretic regime in which case one needs to solve either (\ref{eq:p1})-(\ref{eq:p3}) or (\ref{eq:p21})-(\ref{eq:p23}). 
We illustrate this regime in figure~\ref{fig:r} where the amplitude of oscillation has been obtained from solving (\ref{eq:p1})-(\ref{eq:p3}) numerically. 
The small inset of the figure illustrates the shape of the phase response across the resonance for this family of curves. 
The phase response for a larger set of voltages is shown in figure~\ref{fig:phi}.

\begin{figure}[t]
\begin{center}
\epsfig{file=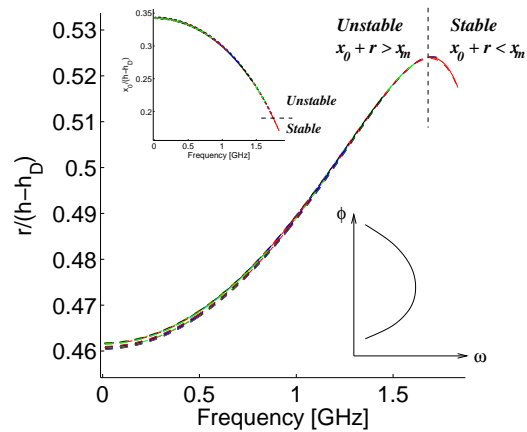, width=8cm,clip} 
\end{center}
\caption{Large amplitude solutions in hysteretic regime ($V_+=0.09,...,0.17$ V). Most of the solutions have oscillation amplitudes too large to be stable. Note, however, that perturbation theory predicts a stable region for the bias $V_+=0.17$ V (red curve). (\full) stable solutions, (\dashed) unstable solutions. \label{fig:r2} }
\end{figure}

In figure~\ref{fig:phi} full lines correspond to stable solutions and dashed lines to unstable solutions. We note that there exists a set of low frequency solutions to the full equations Eqs.~(\ref{eq:p1})-(\ref{eq:p3}) for the intermediate voltages ($V_+=0.09,...,0.17 $ V). These solutions cannot be described by the frequency response equation (\ref{eq:fresp}) and are large amplitude solutions. The corresponding amplitudes and average displacements for these solutions are shown in figure~\ref{fig:r2}. They are mostly unstable due to the fact that their peak oscillation amplitude exceeds $x_m$. Note, however, (see figure~\ref{fig:phi} and figure~\ref{fig:r2}) that perturbation theory predicts 
that some of these solutions have stable ($V_+=0.17$ V) regions. These regions cannot be reached by sweeping the frequency down from a higher frequency but appears as a disconnected manifold of stable solutions.

\subsection{Large driving, large hysteresis}
As the driving increases even more the peak amplitude of the oscillator will eventually reach the point where the curvature of $V_{\rm eff}$ changes from positive to negative.
This is clearly seen in the phase response in figure~\ref{fig:phi} as a gap opens up around $\phi=\pi/2$.
The corresponding large amplitude vibrations in this regime are rendered unstable when the oscillator amplitude reaches the point where it ``rolls over the hill'' 
($x_0+r>x_m$) and the system snaps into contact.The oscillation amplitude and average displacement for one such solution $V_+=0.23$ V are shown as the full lines in figure~\ref{fig:anal_num}.

\section{Validity of perturbation theory and output power}
\begin{figure}[t]
\begin{center}
\epsfig{file=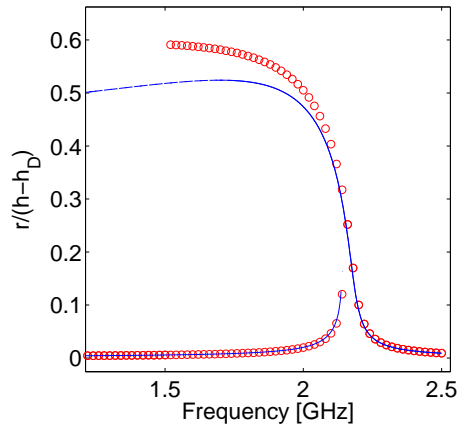,width=8cm,clip}
\end{center}
\caption{Oscillation amplitude $r$ and average displacement $x_0$ in the ``phase gapped'' oscillation regime. Solid lines represent result from perturbation theory for a bias of $V_+=0.23V$ while circles (\opencircle) mark the result of numerical integration of the full system of ODE:s. For small amplitudes the agreement between numerical integration and perturbation theory is very good while perturbation starts to fail for larger amplitude due to the anharmonicity of the tip motion.\label{fig:anal_num}}
\end{figure}

We have also compared the analytical expressions with direct numerical integration of the differential equations. For small amplitude vibrations, including the primary hysteretic regime we find excellent agreement between numerical simulations and perturbation theory. Deviations from the predictions of perturbation theory are only seen for the large amplitude vibrations in the regime where there is a ``phase gap''. An example of such a comparison is shown in figure~\ref{fig:anal_num} where we clearly see the good agreement for small amplitude vibrations and the deviations at larger amplitudes. By looking at the detailed motion of the tube tip at large amplitudes one can see that the approximation by a pure harmonic motion is no longer a good approximation and higher harmonics have to be taken into account.  

Furthermore, perturbation theory predicts the existence of disconnected manifolds of stable large amplitude orbits that cannot be reached by sweeping down in frequency. 
We have not been able to detect any such orbits in the numerical simulations. This may be either due to the solutions being unphysical solutions to the perturbation theory equations or that appropriate initial conditions have not been used in the numerical simulations.  

\begin{figure}[t]
\begin{center}
\epsfig{file=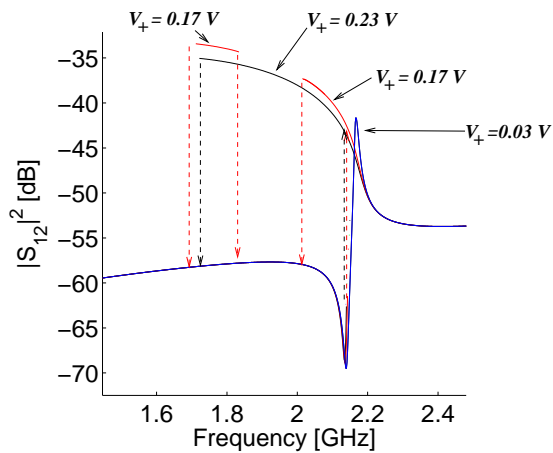,width=9cm,clip}
\end{center}
\caption{$|S_{21}|^2 (\equiv P_{\rm in}/P_{\rm out})$ as a function of frequency for transmission line impedance of $Z_0=10k\Omega$, or 200 devices in parallel connected to common 50 $\Omega$ transmission lines (see text). Note the disconnected stable manifold at $V_+=0.17 V$ and $f\approx 1.8$ GHz.\label{fig:S21}}
\end{figure}

The above only concerned matters related to the averaging method. Another issue is related to the assumption of large impedance mismatch (smallness of $\epsilon$).
Typical transmission line impedances are of the order $50$ $\Omega$. With capacitances in the aF range this means that we have $\epsilon < 10^{-6}$ which makes the approximation excellent. On the other hand this also means that transmitted power is very low. The obvious way to remedy this is to connect several devices in parallel, coupled to common gate and drain electrodes. For an array of $N$ devices coupled this way one finds that $Z_0$ should be replaced by an effective transmission line impedance of $Z_0^{\rm eff}=NZ_0$. Shown in figure~\ref{fig:S21} is the $S_{21}$-parameter for such an effective impedance of $10 k\Omega$ calculated using (\ref{eq:S21}) for biases corresponding to the three different oscillation regimes. The arrows denotes the associated jumps in the hysteresis curve. 

Comparisons with numerical integration show that treating the system to lowest order in $\epsilon$ as we have done here is a good approximation as long as $|S_{21}|\ll 1$.
As $|S_{21}|\sim 1$ the effect of dissipation in the tube-source contact and power delivered to the gate/drain result in a loaded Q-factor exceeding the bare mechanical Q-factor leading to a broadening of the resonances.

\section{Conclusions}
We have carried out an investigation of the RF-response in a three-terminal carbon nanotube resonator structure. By employing perturbation theory we have reduced the 
problem of determining and classifying the frequency response to that of solving a set of algebraic equations. We have found three distinct oscillatory regimes: linear response, hysteretic and a ``phase gapped'' regime which can lead to large hysteresis loops in the frequency response. Comparisons with direct numerical integration have shown that 
perturbation theory is qualitatively correct, and quantitatively  accurate as long as the amplitude of oscillation is not too large. 
The perturbative treatment includes also terms related to parametric driving. We find no qualitative change in the behavior in the RF-response that arises from incorporating such terms.  

\ack
This project has been supported by Nokia Research Center (A. I.), Swedish Foundation for Strategic Research (SSF) (J. M. K.), EU through the Nano-RF project FP6-2005-028158 (J. M. K.). We are grateful for stimulation discussions with Eleanor Campbell, Anders Eriksson, Sang-Wook Lee and Jukka Wallinheimo.

\appendix

\section{Derivation of (\ref{eq:lumpad})}
Here we outline the derivation of the lumped model for tube vibrations used in the paper.
We begin by writing the solution to the PDE in the form
\begin{equation*}
u(x,t)=\Delta_{f_0}(x)+\sum_{n=0}^{\infty}\gamma_n(t)u_n(x).
\end{equation*}
Here $u_n(x)$ are the eigenmodes with frequency $\Omega_n$ satisfying the homogeneous equation $(-\Omega_n^2+({EI}/{\rho A})\partial_x^4)u_n(x)=0$
while the offset $\Delta_{f_0}(x)$ satisfies $({EI}/{\rho A})\partial_x^4\Delta_{f_0}(x)=f_0(x)$
for some function $f_0(x)$ to be determined (typically this is the non-vanishing part of the time averaged force).
Inserting the solution into the PDE for $u(x,t)$ 
and projecting onto a normal mode $u_m(x)$ we find 
\begin{eqnarray}
\fl \ddot{\gamma}_m(t)+\omega_m^2\gamma_m(t)=-L^{-3}\int_{0}^{L}dx u^\dagger_m(x)f_0(x)+\frac{1}{2\rho AL^3}\sum_{i={\rm G,D}}[\Delta V_i(t)]^2\int_{0}^{L}dx u^\dagger_m(x)\frac{\delta C_i[u,t]}{\delta u(x)}.\nonumber
\end{eqnarray}
We use here the normalization convention $\int_0^L dx u_m^\dagger(x) u_n(x)=L^3\delta_{nm}$.
We will typically work with a system where the driving only excites resonances of the fundamental mode  $u_0(x)$. So we can
quite safely make the approximation $u(x,t)=\Delta(x)+\gamma_0(t)u_0(x)$.
The differential equation then reads 
\begin{eqnarray}
\fl\ddot{\gamma}_0(t)+\Omega_0^2\gamma_0(t)=-L^{-3}\int_{0}^{L}dx u^\dagger_0(x)f_0(x)+\sum_{i={\rm G,D}}\frac{[\Delta V_i(t)]^2}{2\rho AL^3}\int_{0}^{L}dx u^\dagger_0\frac{\delta C_i[\Delta(x)+\gamma_0(t)u_0(x)]}{\delta u(x)}.\nonumber
\end{eqnarray}
We will write this in terms of the displacement at the tip of the cantilever 
$x_{\rm T}(t)=\Delta(L)+\gamma_0(t)u_0(L)$,
\begin{eqnarray}
\fl{\ddot{x}_{\rm T}(t)+\Omega_0^2x_{\rm T}(t)}=-\frac{u_0(L)\Omega_0^2}{L^3}\int_{0}^{L}dx u^\dagger_0(x)\epsilon(x)+\Omega_0^2\epsilon(L)\nonumber\\
+\frac{u_0(L)}{2\rho AL^3}\sum_{i={\rm G,D}}\Delta V_i(t)^2\int_{0}^{L}dx u^\dagger_0(x)\frac{\delta C_i[\frac{x_{\rm T}(t)}{u_0(L)}u_0(x)+\epsilon(x)]}{\delta u(x)}.\nonumber
\end{eqnarray}
Here, the static bending has been expressed in terms of the deviation from the shape of the deflection in the lowest mode.
$\Delta(x)=u_0(x){\Delta(L)}/{u_0(L)}+\epsilon(x)$.
Assuming that the statically deformed shape closely resembles the shape of the resonance shape
i.e., we set $\epsilon(x)=0$, we arrive at 
\begin{eqnarray}
\ddot{x}_{\rm T}(t)+\Omega_0^2x_{\rm T}(t)=\frac{u_0(L)}{2\rho AL^3}\sum_{i={\rm G,D}}[\Delta V_i(t)]^2\int_{0}^{L}dx u^\dagger_0(x)\frac{\delta C_i[{x_{\rm T}(t)}\frac{u_0(x)}{u_0(L)}]}{\delta u(x)}.\nonumber
\end{eqnarray}
Finally, using that $u_0^\dagger(x)=u_0(x)$ one finds
\begin{eqnarray}
\ddot{x}_{\rm T}(t)+\Omega_0^2x_{\rm T}(t)&=&\frac{1}{2M_{\rm eff}}\sum_{i={\rm G,D}}[\Delta V_i(t)]^2C_i'(x_{\rm T})\nonumber\\
\label{eq:lump}
\end{eqnarray}
where $M_{\rm eff}\equiv M {L^2}/{u_0(L)^2}\approx M/5.684$ and 
 $\Omega_0={3.516}{L^{-2}}\sqrt{{EI}/{\rho A}}$.

\end{document}